\documentclass[aps,prd,a4paper,floats,nofootinbib]{revtex4} 
\pdfoutput=1
\usepackage{amsmath}
\usepackage{amssymb}
\usepackage{amsfonts}
\usepackage{graphicx, bm}
\usepackage{bm}
\usepackage[usenames]{color}

\newcommand{\rd}{}%\textcolor{red}}

\newcommand{\dnv}{\Delta^{\bm{v}_o}_N}

\newcommand{\dd}{{\cal D}}
\newcommand{\bn}{{\bm{n}}}
\newcommand{\bv}{{\bm{v}_o}}
\newcommand{\bvh}{{\hat{\bm{v}}_o}}
\newcommand{\la}{{\big\langle}}
\newcommand{\ra}{{\big\rangle}}

\def\be{\begin{equation}}
\def\ee{\end{equation}}
\def\ba{\begin{eqnarray}}
\def\ea{\end{eqnarray}}

\begin{document}

\title{The kinematic dipole in galaxy redshift surveys}

\author{Roy Maartens$^{a,b}$, Chris Clarkson$^{c,a,d}$, Song Chen$^a$\\~
\\
\emph{ $^a$Department of Physics \& Astronomy, University of the Western Cape,
Cape Town 7535, South Africa \\
$^b$Institute of Cosmology \& Gravitation, University of Portsmouth, Portsmouth PO1 3FX, United Kingdom\\
$^c$School of Physics \& Astronomy, Queen Mary University of London, London E1 4NS, United Kingdom\\
$^d$Department of Mathematics \& Applied Mathematics, University of Cape Town, Cape Town 7701, South Africa}~\\~\\}

\date{\today}

\begin{abstract}
In the concordance model of the Universe, the matter distribution -- as observed in galaxy number counts or the intensity of line emission (such as the 21cm line of neutral hydrogen)  -- should have a kinematic dipole due to the Sun's motion relative to the CMB rest-frame. This dipole should be aligned with the kinematic dipole in the CMB temperature.  Accurate measurement of the direction  of the matter dipole will become possible with future galaxy surveys, and this will be a critical test of the foundations of the concordance model. The amplitude of the matter dipole is also a potential  cosmological probe. We derive formulas for the amplitude of the kinematic dipole in galaxy redshift and intensity mapping surveys, taking into account the Doppler, aberration and other relativistic effects. The amplitude of the matter dipole can be significantly larger than that of the CMB dipole. Its redshift dependence encodes information on the evolution of the Universe and on the tracers, and we discuss possible ways to determine the amplitude.

\end{abstract}
\maketitle

\section{Introduction}

The concordance model of cosmology assumes statistical isotropy and homogeneity of the Universe. Since the Solar System has a nonzero velocity $\bm{v}_o$ relative to the cosmic frame defined by the cosmic microwave background (CMB), this assumption requires that the matter distribution must have a kinematic dipole which is aligned with that of the CMB\footnote{Provided that the matter kinematic dipole is measured with high enough median redshift to beat down the influence of the local structure dipole (see \cite{Gibelyou:2012ri} for a detailed discussion).}. This is  a critical test of the foundations of the concordance model, as first pointed out and attempted in \cite{eb84}.

Up to now, attempts to measure the kinematic direction $\hat{\bm{v}}_o$ via the matter dipole are compromised by low median redshifts and low number densities (see \cite{Tiwari:2015tba,Bengaly:2016amk,Colin:2017juj,Bengaly:2017zlo} for recent work).
Accurate measurement of $\hat{\bm{v}}_o$ from the matter distribution will become possible with next-generation galaxy surveys that will cover much of the sky out to high redshift and with high number densities \cite{Schwarz:2015pqa,Yoon:2015lta,Ghosh:2016tbj}. 

In addition to the direction $\hat{\bm{v}}_o$, observations measure the amplitude of the matter dipole. This amplitude is determined not only by Doppler and aberration effects, but also by properties of the galaxy survey and by the evolution of the Universe. Here we focus on the amplitude of the kinematic dipole in galaxy redshift surveys, based on number counts or on intensity mapping of HI (or other line emission). This redshift-dependent kinematic dipole has not previously been presented, and measurements of the kinematic dipole up to now have used the projected, two-dimensional number counts.

We analyse the contributions to the matter kinematic dipole, using observational variables and taking care to include all relativistic effects. We start from the basic relativistic boost contribution to the number counts, and then include contributions from the expansion history of the Universe, the evolution of source number density and the magnification bias that arises from the magnitude limit of a galaxy redshift survey. Then we treat the case of HI intensity mapping. The redshift-dependent  amplitude of the matter kinematic dipole encodes independent information about properties of the Universe and the tracer.
We compare the amplitude of the matter kinematic dipole in the angular correlation function to that of the CMB kinematic dipole. We also discuss possible ways to determine the amplitude of the matter kinematic dipole. 

\section{Kinematic dipole in galaxy number counts}

Assuming that the observer moves with velocity $\bm{v}_o$ as measured by CMB observations 
(with\footnote{We choose units with $c=1$ so that $v_o$ is dimensionless.}
 $v_o\approx 10^{-3}$), what is the resulting dipole in number counts for a galaxy redshift survey? We break the answer down into steps.

\subsection{Doppler and aberration effects}

The boosted observer's 4-velocity $\tilde{u}^\mu_o$ is related to the CMB rest-frame observer's 4-velocity $u^\mu_o$ by
\be\label{tu}
\tilde{u}^\mu_o=\gamma(v_o)\big[u^\mu_o+v^\mu_o\big]=u^\mu_o+v^\mu_o+O(v_o^2) \qquad\mbox{where}\qquad u_o^\mu v_{o\mu}=0.
\ee
From now on, we neglect terms of $O(v_o^2)$ and second-order terms in metric perturbations. 
In the CMB rest-frame, $v^\mu_o=(0,\bm{v}_o)$. The boosted observer measures a different cosmological redshift $\tilde z$ to that measured in the CMB frame, $z$. For a source in the unit direction $n^\mu=(0,\bm{n})$ in the CMB frame, we have $u^\mu_o n_\mu=0$, and the backward lightray vector is $k_o^\mu \propto -u_o^\mu+n^\mu$. It follows that 
\be
{1+\tilde z \over 1+z}={(u_\mu k^\mu)_o \over (\tilde u_\mu k^\mu)_o }={1\over 1+\bm{n}\cdot\bm{v}_o},
\ee 
since the contributions to $z,\tilde z$ at the source are the same and cancel. Thus the observed redshifts are Doppler related: 
\be \label{zt}
1+\tilde z=(1+z)\big(1-\bm{n}\cdot\bm{v}_o\big).
\ee

The direction of sources is affected by aberration, $\bm{n} \to \tilde{\bm{n}}$. The photon wave vector $k^\mu$ is invariant under a boost, so that $-\tilde{u}_o^\mu+\tilde{n}^\mu=(1-\bm{n}\cdot\bm{v}_o)[-u^\mu_o+n^\mu]$, where we used \eqref{zt}. Projecting into the rest-space and using  \eqref{tu}, we obtain
\ba
\tilde{\bm{n}}=\big(1-\bm{n}\cdot\bm{v}_o \big)\bm{n}+\bm{v}_o .
\label{tn2}
\ea
Taking the cross product with $\hat{\bm{v}}_o$, we have
\be\label{tn3}
% \cos\tilde\theta= \big(1-\bm{n}\cdot\bm{v}_o \big)\cos\theta + v_o,\qquad 
 \sin\tilde\theta=\big(1-\bm{n}\cdot\bm{v}_o \big)\sin\theta,
\ee
where $\theta={\rm arccos}\,(\bm{n}\cdot \hat{\bm{v}}_o)$ and similarly for $\tilde\theta$. 
By writing $\tilde\theta=\theta+\delta\theta$ and expanding \eqref{tn3}, we find
\be\label{tn4}
\tilde\theta=\theta-v_o\sin\theta.
\ee
This shows that source images are aberrated towards the forward dipole direction. 
 The azimuthal angle is unaffected by boosting, so that solid angles at the boosted and rest-frame observers are related by
\be\label{domt}
 {d\tilde\Omega_o\over d\Omega_o}= {\sin\tilde\theta\over\sin\theta}{ d\tilde\theta\over d\theta} 
 {d\varphi\over d\varphi}=1-2\,\bm{n}\cdot\bm{v}_o,
\ee
where we used \eqref{tn3} and \eqref{tn4}.
Since the intrinsic area of a source is invariant, we have $\tilde d_A^{\,2}\,d\tilde\Omega_o=d_A^2\,d\Omega_o$, and hence the angular diameter distance obeys
\be\label{dat}
\tilde d_A(\tilde z,\tilde\bn)=d_A(z,\bn)\big[1+\bn\cdot\bv\big].
\ee
The reduction of the solid angle along the forward dipole direction corresponds to an increase in the area distance, since the object of given intrinsic size appears smaller and hence further away. 

The total number of galaxies is invariant:
\be
\tilde N \,d\tilde z\,d\tilde\Omega_o= N\,dz\,d\Omega_0,
\ee
where $N, \tilde N$ are the numbers observed per redshift interval and per solid angle. Then, using \eqref{zt} and \eqref{domt}, the number count for the boosted observer is given by
\be\label{nd}
\tilde N(\tilde z,\tilde\bn)=N(z,\bn)\big[1+3\,\bm{n}\cdot\bm{v}_o\big].
\ee

The kinematic dipole in the number counts is thus $3\,\bm{n}\cdot\bm{v}_o$, which  can be thought of as the special relativistic approximation to the matter dipole. There are two corrections that are needed: a redshift correction and a magnitude correction. We deal with these in turn.

\subsection{The redshift correction}

Equation \eqref{nd} contains implicitly an important correction arising from the difference between the observed ($z$ and $\tilde z$) and background ($\bar z$) redshifts. This becomes apparent when we consider the fractional fluctuations in the \rd{observed} number counts (at first order in perturbations):
\be 
\Delta_{\tilde N}(\tilde z,\tilde\bn)={\tilde N(\tilde z,\tilde\bn)-\bar N(\tilde z) \over \bar N(\tilde z)}
={N(z,\bn)\big[1+3\,\bm{n}\cdot\bm{v}_o\big]-\big[\bar N({z})-(1+{z})\bm{n}\cdot\bm{v}_o\,d\bar N({z})/d{z} \big]\over
\bar N({z})-(1+{z})\bm{n}\cdot\bm{v}_o\,d\bar N({z})/d{z}},
\ee
where we used \eqref{zt} and we assume that the survey detects all galaxies, without any magnitude limit; the case of a magnitude limited survey is considered below.  Thus 
\be\label{okd1}
\Delta_{\tilde N}(\tilde z,\tilde{\bn})=\Delta_N(z,\bn)+\left[ 3+{d\ln \bar N({z})\over d\ln(1+{z})} \right] \bm{n}\cdot\bm{v}_o
=\Delta_N({\bar z},\bm{n})+\left[ 3+{d\ln \bar N(\bar z)\over d\ln(1+\bar z)} \right] \bm{n}\cdot\bm{v}_o ,
\ee
{where the second equality follows since we are working to first order.} \rd{Note that $\Delta_N$ includes redshift and volume perturbations: see \eqref{obsn} below.}
The contribution of $\tilde\bn-\bn$ to $\Delta_{\tilde N}$ is second order in perturbations and we can rewrite \eqref{okd1} as
\be\label{okd}
\Delta_{\tilde N}(\tilde z,{\bn})=\Delta_N({\bar z},\bm{n})+\dnv({\bar z},\bn),\qquad \dnv({\bar z},\bn)=\left[ 3+{d\ln \bar N(\bar z)\over d\ln(1+\bar z)} \right] \bm{n}\cdot\bm{v}_o.
\ee

This shows how the evolution of $\bar N$ modifies the kinematic dipole.  
The observed proper volume element in the background is $dV=dR\,dA=(adr)\,(\bar {d}_A^{\,2}\,d\Omega_o)= (a/H)d\bar z\, (ar)^2d\Omega_o$, where $r$ is the line-of-sight comoving distance. Then the
 background number of sources per $\bar z$ per $\Omega_o$ is given by
\be \label{bnd}
\bar N=\left({r^2 \over H } \right)\big[(1+\bar z)^{-3}\bar n_s \big],
\ee
where $\bar n_s$ is the proper number density of sources. The term in round brackets is the comoving volume per redshift per solid angle, while the term in square brackets is the comoving number density. Thus the redshift correction in \eqref{okd} encodes a volume and a number density contribution.

Using \eqref{bnd} in \eqref{okd}, we can display the 3 types of contribution to the kinematic dipole:  
\be \label{dip1}
\dnv=\left\{ 3+\left[{\dot H \over H^2} +{2(1+\bar z)\over r H}\right] +{d\ln \big[(1+\bar z)^{-3}\bar n_s \big]\over d\ln(1+\bar z)}\right\} \bm{n}\cdot\bm{v}_o.
\ee
The first term in braces is the special relativistic Doppler + aberration contribution; the second term, in square brackets, is the volume part of the redshift correction;
 the final term is the number density part of the redshift correction. This last term is given by the `evolution bias' of the tracer \cite{Challinor:2011bk,Jeong:2011as}:
\be\label{eb}
b_e=-{d\ln\big[(1+\bar z)^{-3}\bar n_s \big]\over d\ln(1+\bar z)},
\ee
which measures the deviation from the idealized case of conservation of comoving number density, $\bar n_s\propto a^{-3}$, for which $b_e=0$. {From now on we drop the overbar on $z$ in first-order expressions, since the difference is second order.}

In summary, the kinematic dipole in the number count contrast, when we assume that there is no magnitude limit, is given by
\ba \label{dip2}
\dnv(z,\bn) = \dd(z)\, \bm{n}\cdot \bvh ~~\mbox{where}~~\dd=\left[3+{\dot H \over H^2} +{2(1+z)\over rH}-b_e \right]v_o.
\ea
Here and below we drop the overbar on the background redshift where there is no ambiguity.
Equation \eqref{dip2} holds for any perturbed Friedmann spacetime. For the concordance model, we have
\ba\label{dipc}
\dd(z)= \left[3-{3\over2}\Omega_m(z)+{2(1+z)\over r(z) H(z)}-b_e(z)\right]v_o \qquad\quad \mbox{LCDM.}
\ea
At low redshifts, the term $2(1+z)/rH$ dominates: e.g., for $z=0.1$, it is $\sim 20$. 

\subsection{Including the magnitude limit}

Now we take account of the fact that  only sources with observed magnitude $m<m_*$ are detected. Fluctuations in the observed magnitude, which can be given by fluctuations in the luminosity distance $d_L$, affect the number counts, so that $\Delta_{\tilde N}(\tilde z,\bn)$ is replaced by  \cite{Challinor:2011bk,Jeong:2011as,Alonso:2015uua}
\be\label{m*}
\Delta_{\tilde N}(\tilde z,\bn, m<m_*)= \Delta_{\tilde N}(\tilde z,\bn)- {5}s(\tilde z,m_*)\, {\Delta d_L(\tilde z,\bn) \over \bar d_L(\tilde z)}.
\ee
Here,
\be
s(\bar z,m_*)={\partial \log \bar {\cal N}_s(\bar z,m<m_*) \over \partial m_*}
\ee
is the magnification bias, where $\bar {\cal N}_s$ is the cumulative luminosity function:
\be
\bar {\cal N}_s(\bar z,m<m_*)={2\over 5}\ln 10\int_{-\infty}^{m_*}dm\, \bar n_s(\bar z,m).
\ee

We are interested in the kinematic dipole of \eqref{m*}, which requires the kinematic dipole of the luminosity distance fractional fluctuation. Using the general reciprocity relation $\tilde{d}_L=(1+\tilde z)^2\tilde d_A$, together with \eqref{zt} and \eqref{dat}, we find that
\be\label{dlt}
\tilde d_L(\tilde z,\bn)=d_L(z,\bn)\big[1-\bn\cdot\bv\big].
\ee
The luminosity distance in the direction of the boost is decreased since the observed flux $L/(4\pi \tilde d_L^2)$ increases while the intrinsic luminosity $L$ is invariant. 

Taking account of the difference between observed and background redshifts, we find from \eqref{zt} and \eqref{dlt} that the fractional fluctuation is given by
\be\label{ddl}
\Delta_{ \tilde d_L}(\tilde z, {\bn})=\Delta_{d_L}(z,\bm{n})+\left[ -1+{d\ln \bar d_L({z})\over d\ln(1+{z})} \right] \bm{n}\cdot\bm{v}_o.
\ee
In the background, $\bar d_L=(1+\bar z)r$, so that the kinematic dipole in the fractional luminosity distance fluctuation is\footnote{This is in agreement with \cite{Bonvin:2005ps}, noting that their $\bn$ is minus ours. Note also that the dipole for the area distance is the same as \eqref{dld}.\\}
\be\label{dld}
\Delta_{d_L}^{\bv}(z,\bm{n})={(1+z)\over r(z)H(z)}\,\bn\cdot\bv.
\ee

Finally, \eqref{m*} and \eqref{dld} lead to:
\be\label{dipm}
\dnv(z,\bn,m<m_*)=\dnv(z,\bn)-{5(1+z)s(z,m_*)\over r(z)H(z)}\, \bm{n}\cdot\bm{v}_o=
{\cal D}_{\rm gal}(z,m_*)\, \bm{n}\cdot \bvh,
\ee
where the amplitude of the dipole is
\ba \label{dipm2}
{\cal D}_{\rm gal}= \left[3+{\dot H \over H^2}+(2-5s){(1+z)\over rH}-b_e\right]v_o.
\ea
Since the number density is a function of $m$,  \eqref{eb} should now be replaced by 
\be\label{eb2}
b_e(\bar z,m_*)=-{\partial\ln\big[(1+\bar z)^{-3}\bar {\cal N}_s(\bar z,m<m_*)\big]\over \partial\ln(1+\bar z)}.
\ee

Equation \eqref{dipm2}
generalizes \eqref{dip2} and gives the theoretical prediction for the amplitude of the kinematic dipole in the number count contrast, for a magnitude-limited galaxy redshift survey.  

To our knowledge, this expression is not given in previous literature on the cosmic kinematic dipole,\footnote{
${\cal D}_{\rm gal} \bm{n}\cdot \bvh$ is part of the general equation for the number count contrast observed on the past lightcone, but terms at the observer are typically neglected in the final expression (e.g. \cite{Challinor:2011bk,Jeong:2011as,Hall:2012wd}). There is also a monopole in the number count contrast, determined by the metric gravitational potentials at the observer, but that is not relevant here.} which has focused on the projected two-dimensional number counts.
%In \cite{Bertacca:2014hwa} these terms are included, and agree with our independent derivation.}
 It shows how the redshift-dependent amplitude of the kinematic dipole in the matter distribution is modified from the naive special relativistic approximation not only by the expansion of the Universe, but also by the astrophysical characteristics ($b_e$ and $s$) of the tracer used  to map the matter distribution.

\subsection{HI intensity mapping}

The fractional fluctuations of brightness temperature for intensity mapping of the neutral hydrogen 21cm line emission after reionization,\footnote{The dipole in the eopoch of reionization is considered in \cite{Slosar:2016utd}.} 
can be obtained at first order from the number count contrast of HI atoms (see  \cite{Hall:2012wd,Alonso:2015uua} for details). 
Perturbations in the brightness temperature are determined by those in the number count and in the luminosity distance:
$\Delta_{T_{\rm IM}}=\Delta_{N_{\rm IM}}-2\Delta d_L/\bar d_L$. Comparing this with \eqref{m*}, it follows that
the map from number count to temperature contrast is given by using an effective $s=2/5$ in \eqref{dipm2}:
\ba \label{dipi}
\dd_{\rm IM}= \left[3+{\dot H \over H^2}-b_e\right]v_o\qquad\quad  \Big(s_{\rm IM} ={2\over5}\Big),
\ea
where 
\ba
 \label{beh}
b_e=-{d\ln\big[(1+\bar z)^{-3}\bar n_{\rm IM} \big]\over d\ln(1+\bar z)} = 
-{d\ln\big[(1+\bar z)^{-2}H\bar T_{\rm IM} \big]\over d\ln(1+\bar z)}.
\ea

\section{Kinematic dipole in the angular correlation function}

{Galaxy surveys naturally include the kinematic dipole, i.e., observations measure the total fluctuation:} 
\ba
\Delta_{\tilde N}=\Delta_N+\dnv, 
\ea
where $\Delta_{N}$ is the observed fluctuation at the source: 
\be\label{obsn}
\Delta_{N}=\delta_N -{(1+z)\over H}\,\partial_r \big(\bm{v}\cdot\bn\big)+(5s-2)\kappa +\Delta^{\rm uls}.
\ee
The second term on the right is the observational correction from redshift space distortions and the third term is  the  lensing magnification correction. The last term represents the ultra-large scale relativistic observational corrections (see \cite{Challinor:2011bk,Jeong:2011as,Alonso:2015uua}).
The angular correlation function for a redshift survey is defined by
\be
\la\Delta_{\tilde N}(z,\bn,m<m_*)\, \Delta_{\tilde N}(z',\bn',m'<m_*)\ra= \sum_\ell{(2\ell+1)\over 4\pi}\, \tilde C_\ell(z,z',m_*)\, P_\ell(\bn\cdot\bn').
\ee

Since the kinematic dipole term $\dnv$ does not correlate with the terms at the source, $\Delta_N$, we have
\ba
\la\Delta_{\tilde N}(z,\bn,m<m_*)\, \Delta_{\tilde N}(z',\bn',m'<m_*)\ra 
&=& \la\Delta_{N}(z,\bn,m<m_*)\, \Delta_{N}(z',\bn',m'<m_*)\ra 
\nonumber\\
&&{}+\la\dnv(z,\bn,m<m_*)\, \dnv(z',\bn',m'<m_*) \ra
\nonumber\\ \label{cd}
&=& \sum_\ell{(2\ell+1)\over 4\pi}\, C_\ell(z,z',m_*) P_\ell(\bn\cdot\bn') + {3\over 4\pi}C_1^{\bv}(z,z',m_*)\,\bn\cdot\bn'.
~~~~
\ea
The dipole in the boosted frame is related the intrinsic dipole in the rest-frame and the kinematic dipole as: 
\be \label{cd2}
\tilde C_1=C_1+C_1^{\bv}. 
\ee

By \eqref{dipm}, using 
\be
\la (\bn\cdot\bvh)\,(\bn'\cdot\bvh)\ra={1\over3}  \,\bn\cdot\bn', 
\ee
we find that the kinematic dipole is
\be \label{c1}
C_1^{\bv}(z,z',m_*)= {4\pi\over 9} \,{\cal D}(z,m_*)\,{\cal D}(z',m_*),
\ee
where for galaxies ${\cal D} ={\cal D}_{\rm gal}$ is given by \eqref{dipm2}, and for intensity mapping, ${\cal D} ={\cal D}_{\rm IM}$ is given by \eqref{dipi}.

\begin{figure}[h]%[t!]
\includegraphics[width=0.59\columnwidth]{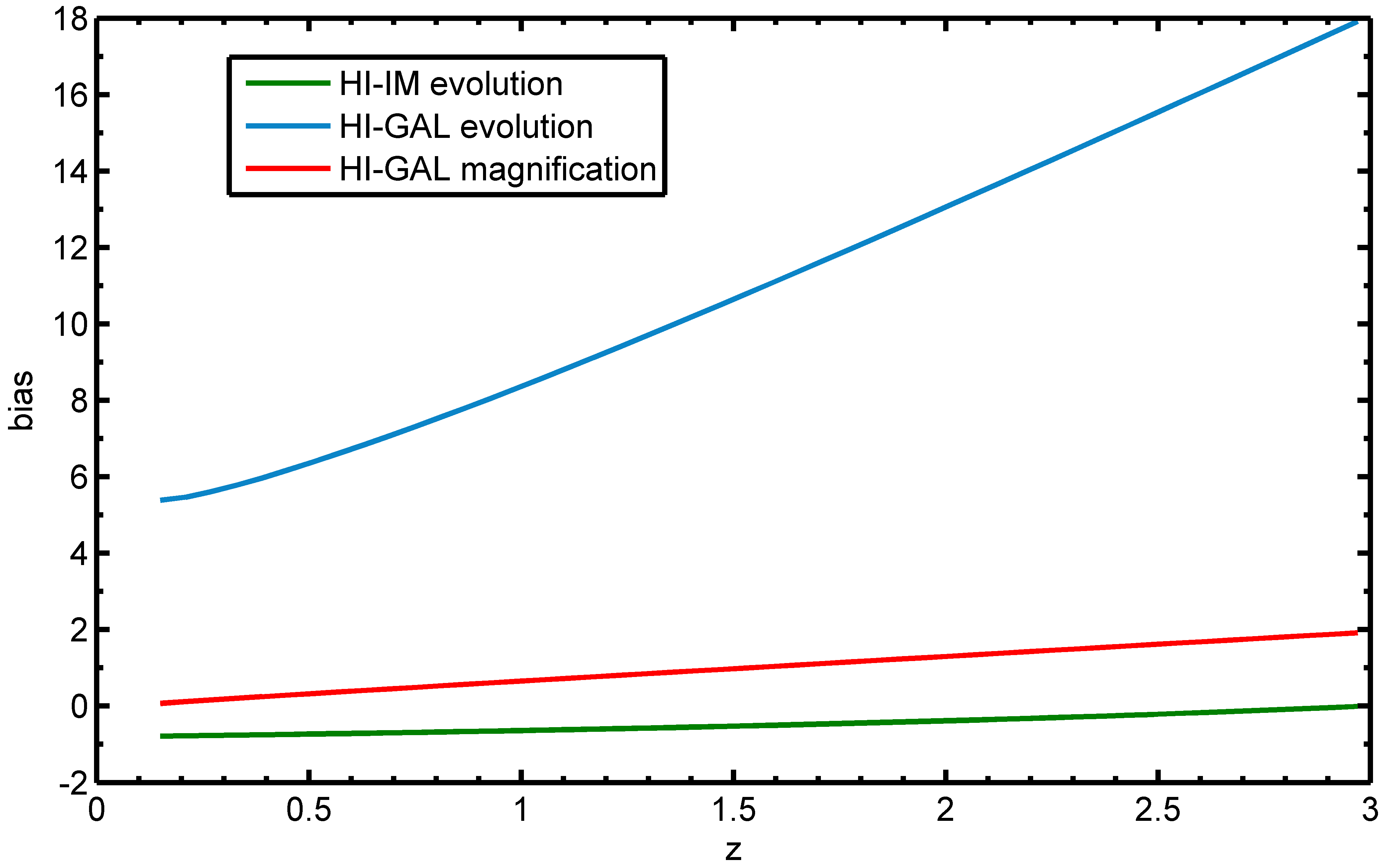} 
\caption{The evolution and magnification biases in SKA galaxy (`gal') and intensity mapping (`IM') surveys. (The effective magnification bias for IM, $s_{\rm IM}=2/5$, is not shown.)}
\label{fbs}
\end{figure}
\begin{figure}[h]%[t!]
\includegraphics[width=0.49\columnwidth]{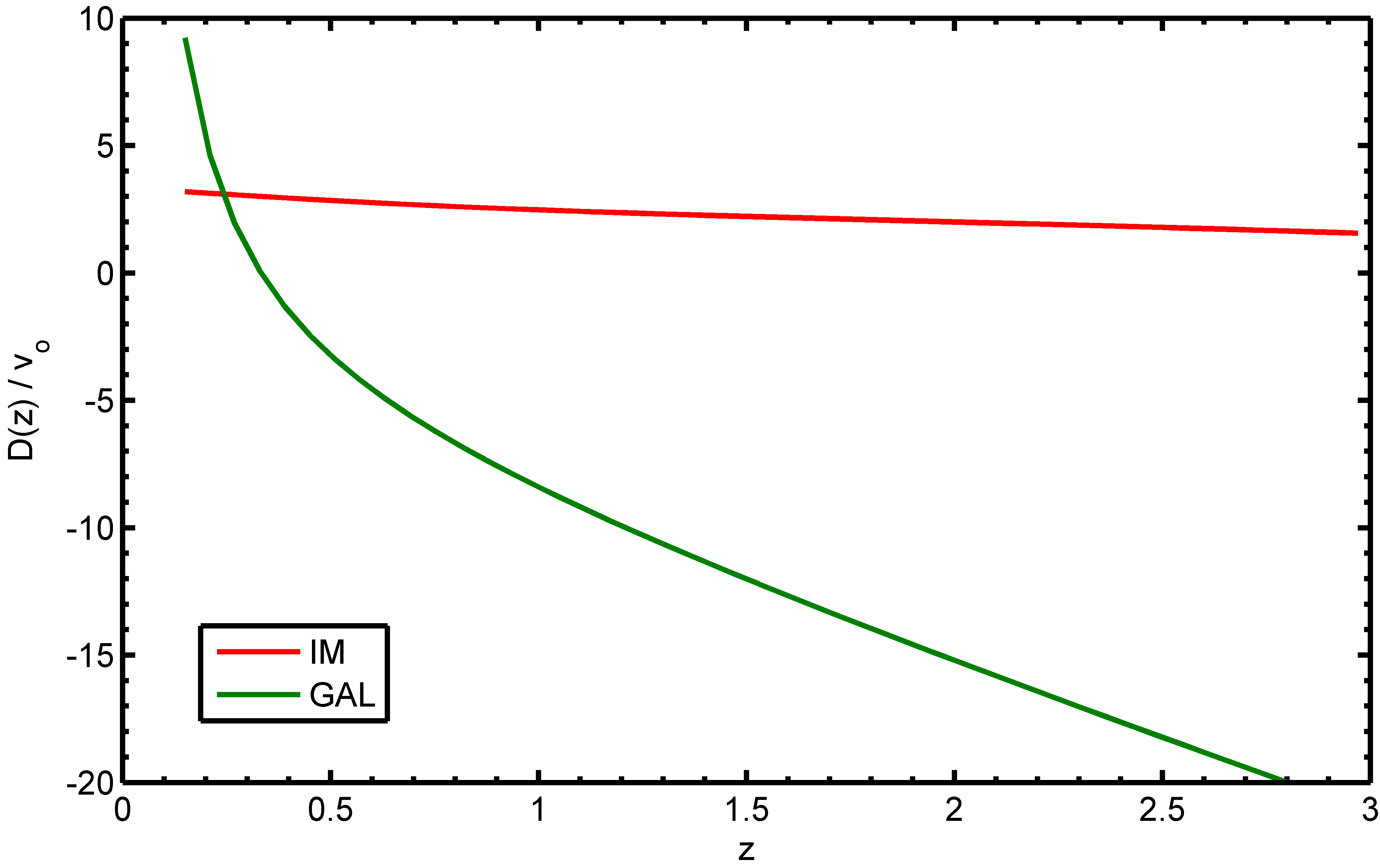}~~
\includegraphics[width=0.49\columnwidth]{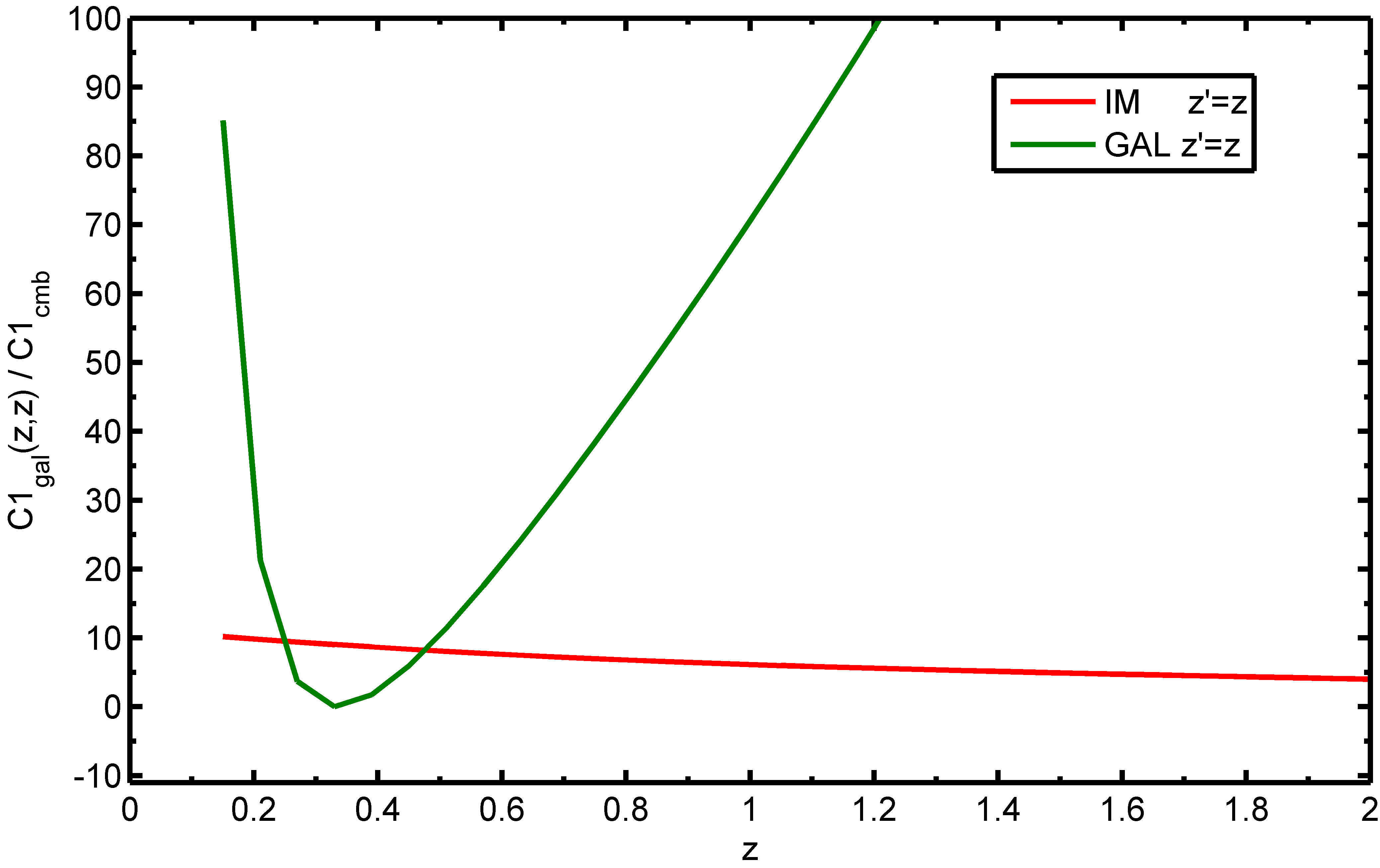}\\~\\~\\
\includegraphics[width=0.49\columnwidth]{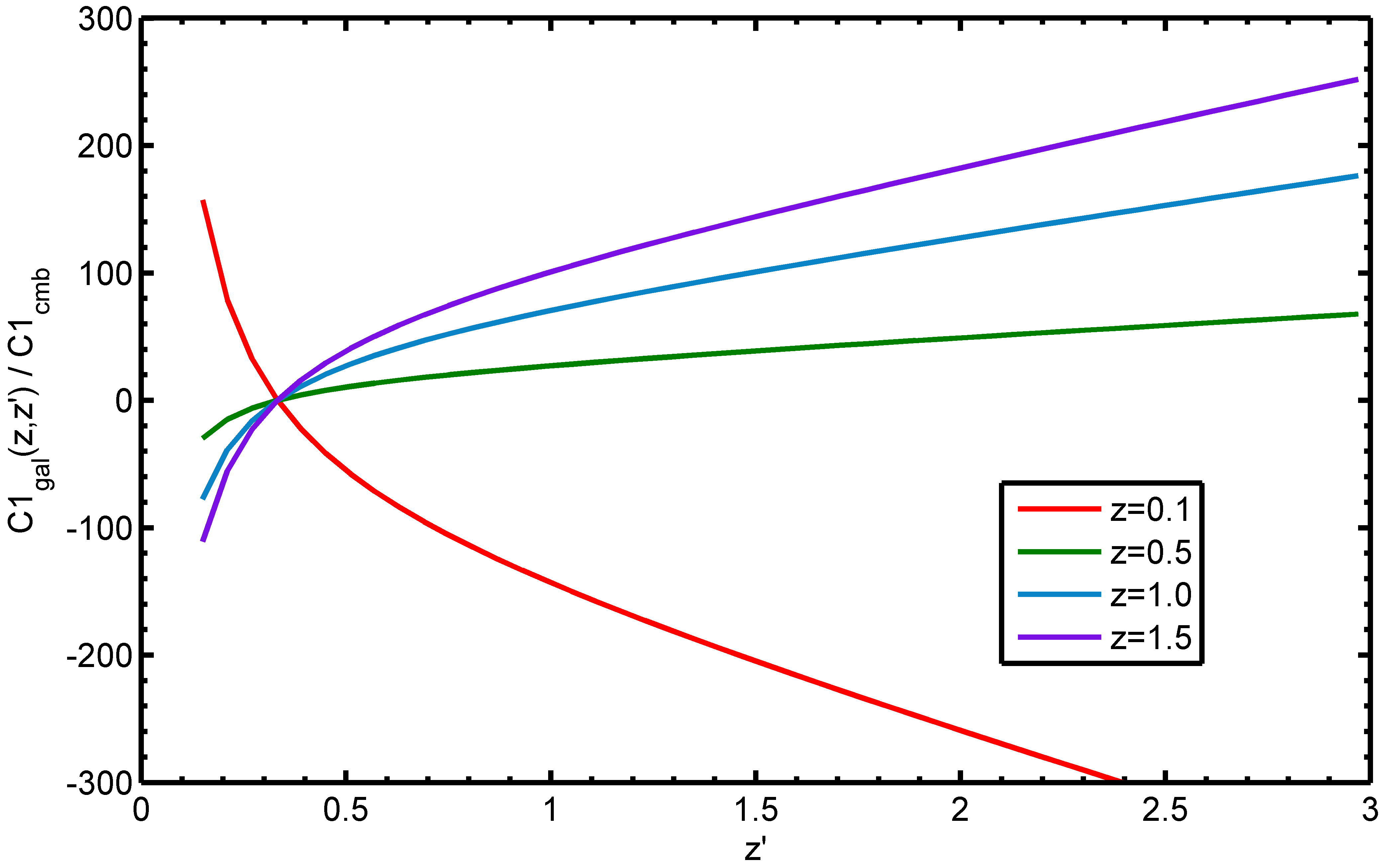}~~ 
\includegraphics[width=0.49\columnwidth]{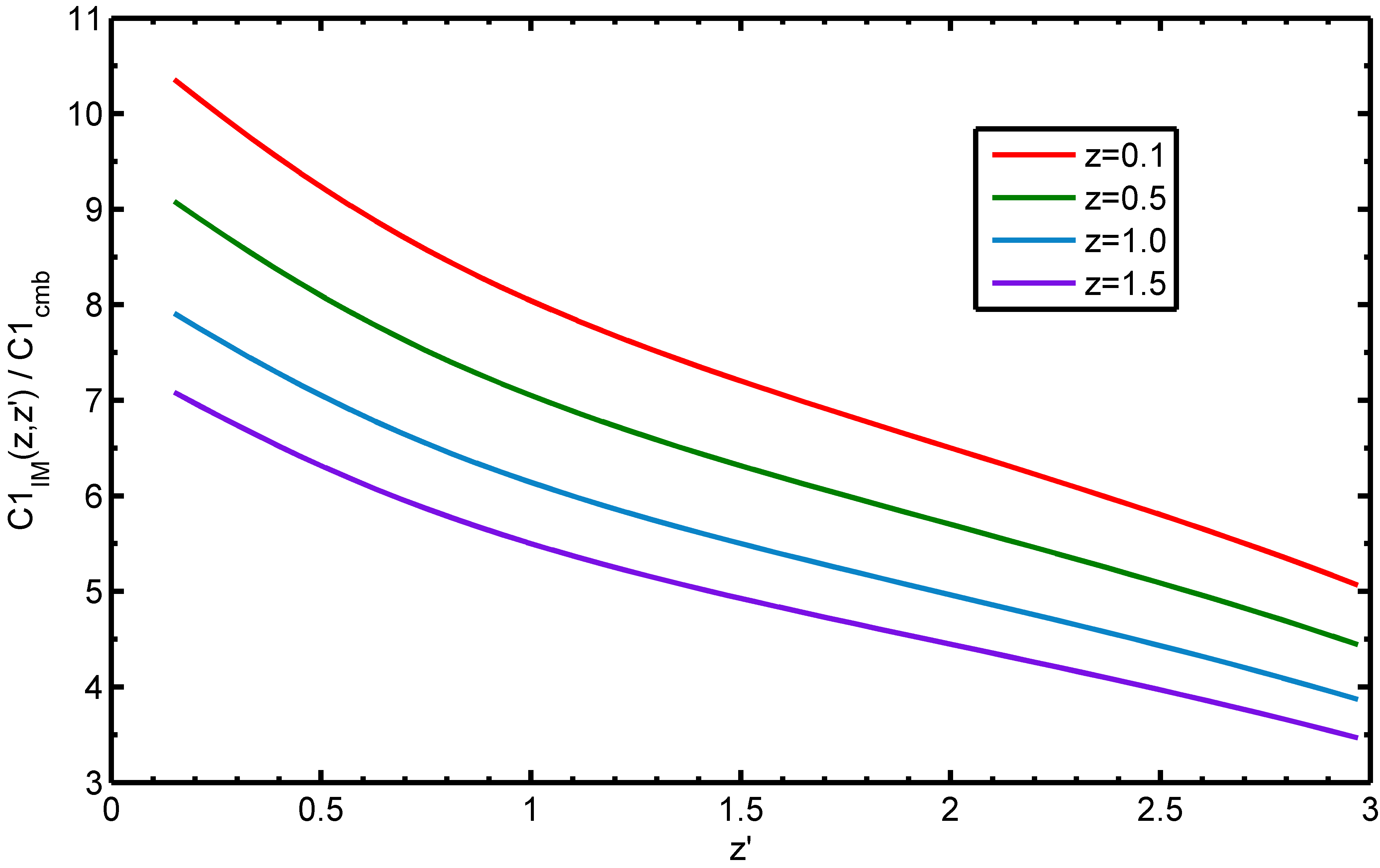}   
\caption{The kinematic matter dipole in SKA galaxy (`gal') and intensity mapping (`IM') surveys. \\
%In all plots the minimum redshift is 0.15.\\ 
{\em Upper:} The dipole in number counts/ intensity {as a fraction of the average, and in units of $v_o$, or equivalently,} relative to the CMB  [see \eqref{rat}] ({\em left}); the dipole in the angular correlations at the same redshift, $z'=z$, {in units of $4\pi v_o^2/9$, or equivalently,} relative to the CMB [see \eqref{kdr}] ({\em right}).\\ 
{\em Lower:} The dipole in the angular correlations at different redshifts, in the galaxy ({\em left}) and intensity mapping ({\em right}) surveys.}
\label{fdc}
\end{figure}

\subsection{{Predicted amplitude of the kinematic dipole}}

In the concordance model, the matter kinematic dipole is aligned with that of the CMB. However the amplitudes of the dipoles are different. 
 The CMB temperature kinematic dipole {as a fraction of the average temperature is   \cite{Aghanim:2013suk}
\be\label{dc}
{\big(\Delta \tilde T\big)^{\bv}\over \bar T} =\dd_{\rm CMB}\, \bn\cdot\bvh \,,\quad \dd_{\rm CMB}=v_o\,.
\ee}
By contrast, the matter dipole is redshift-dependent, and depends on the expansion rate and the astrophysical parameters $b_e$, $s$. 
{The matter kinematic dipole in number counts as a fraction of the average number count, or in brightness temperature as a fraction of the average, follows from \eqref{dipm} or \eqref{dipi}: 
\be \label{dipni}
{\big(\Delta \tilde N\big)^{\bv}\over \bar N} = \dd_{\rm gal}\,\bn\cdot\bvh\,, \qquad
 {\big(\Delta \tilde T_{\rm IM}\big)^{\bv}\over \bar T_{\rm IM}}  =\dd_{\rm IM}\,\bn\cdot\bvh\,.
\ee}

{The amplitude  of the kinematic dipole in counts/ intensity as fraction of the average, 
can conveniently be given in units of $v_o\approx 10^{-3}$. 
By \eqref{dc} and \eqref{dipni}, this is equivalent to the ratio 
\be\label{rat}
\hat{\dd}(z,m_*)\equiv {{\cal D}(z,m_*)\over v_o}={{\cal D}(z,m_*)\over \dd_{\rm CMB}}= 3-{3\over2}\Omega_m(z)+\big[2-5s(z,m_*)\big]{(1+z)\over r(z)H(z)}-b_e(z,m_*),
\ee
where IM corresponds to $s=2/5$. Since $C_{1\,{\rm CMB}}^{\bv}= 4\pi v_o^2/9$, 
it follows from \eqref{rat} that the angular harmonic matter dipole in units of $4\pi v_o^2/9$ is given by 
\be \label{kdr}
\hat{\dd}(z,m_*)\,\hat{\dd}(z',m_*) = {C_1^{\bv}(z,z',m_*)\over 4\pi v_o^2/9}=  {C_1^{\bv}(z,z',m_*)\over C^{\bv}_{1\,{\rm CMB}}}, 
\ee
where we used \eqref{c1}.}

We compute the quantities \eqref{rat} and \eqref{kdr} for two redshift surveys that are planned for the Square Kilometre Array (SKA) \cite{Maartens:2015mra}: an HI galaxy survey in Phase 2, and an HI intensity mapping survey in Phase 1. In order to predict the amplitude of the kinematic dipole in these surveys, we need only the evolution and magnification biases. For the redshift evolution of the Hubble rate, we use the concordance model parameters $\Omega_{m0}=0.3=1-\Omega_{\Lambda 0}, h=0.68$.
\begin{itemize}
\item
{\em Evolution and magnification bias:}\\
Fitting formulas for the evolution and magnification biases of the SKA2 HI galaxy survey are given in \cite{Camera:2014bwa} for a range of possible flux cuts. We choose a flux cut of $3\,\mu$Jy. For the SKA1 HI intensity mapping survey, the effective magnification bias parameter is 2/5 [see \eqref{dipi}]. The brightness temperature is given in \cite{Santos:2015gra}, and then the evolution bias is computed via \eqref{beh} (see also
\cite{Alonso:2015uua}). The results are shown in Fig.~\ref{fbs}.
\item
{\em Amplitude of matter kinematic dipole:}\\
Using the predicted $b_e$ and $s$, we compute the amplitudes \eqref{rat} and \eqref{kdr} for the two SKA surveys. The results are displayed in  Fig.~\ref{fdc}:

The upper left panel shows $\hat{\dd}=\dd/v_o=\dd/\dd_{\rm CMB}$. 

The upper right panel shows $\hat{\dd}^2=C_1^{\bv}/C_{1\,{\rm CMB}}^{\bv}$ for both surveys. 

The lower panels show $\hat{\dd}\hat{\dd}'=C_1^{\bv}(z,z')/C_{1\,{\rm CMB}}^{\bv}$, for the galaxy (left panel) and intensity mapping (right panel) surveys.
\end{itemize}

Figure \ref{fdc} demonstrates that the theoretical amplitude of the matter kinematic dipole {in counts/ intensity has a wide range of possible behaviour}:
\begin{itemize}
\item
The $\hat{\dd}$ curves in the upper left panel explain the behaviour of $\hat\dd^2=C_1^{\bv}/C_{1\,{\rm CMB}}^{\bv}$ in the upper right panel:\\ 
{\em Galaxies:} $\hat{\cal D}$ decreases monotonically and goes through zero at a low redshift, $z\sim 0.3$, so that $\hat\dd^2$ rises after the zero-point.\\
{\em Intensity mapping:} $\hat{\cal D}$ also decreases monotonically but does not go through zero; as a result, $\hat\dd^2$ decreases monotonically.
\item
Similarly, for $\hat{\dd}\hat{\dd}'=C_1^{\bv}(z,z')/C_{1\,{\rm CMB}}^{\bv}$ in the lower panels:\\ 
{\em Galaxies:} $\hat{\dd}\hat{\dd}'$  is negative when $\hat{\cal D}$ has opposite signs at $z$ and $z'$. For the $z=0.1$ curve (red), this occurs when $z'\gtrsim 0.3$, while for the other curves ($z=0.5,1,1.5$), it occurs when $z'\lesssim 0.3$. \\
{\em Intensity mapping:}  $\hat{\dd}\hat{\dd}'$ is always positive since $\hat{\cal D}>0$ at all $z,z'$.
\end{itemize}

The main driver for the difference between the two surveys is evolution bias: in the interval $0.1<z<3$, we have $5< b_e< 19$ for galaxies,\footnote{Note that $b_e$ for galaxies is sensitive to the flux cut. We chose $3\,\mu$Jy -- for higher values (lower sensitivity), $b_e$ is larger.}  
and $-1<b_e<0$ for intensity mapping. The $b_e$ for galaxies is large enough that $-b_e$ in \eqref{dipm2} can produce ${\cal D}_{\rm gal}=0$ at low $z$, and then push  ${\cal D}_{\rm gal}$ to large negative values at high $z$.
Both kinematic dipoles contain the same positive expansion term $3-{3}\Omega_m/2$. The magnification bias contribution is zero for intensity mapping. For galaxies, $0<s<2$, so that the term $(2-5s)(1+z)/ rH$ is positive and large at low $z$, but negative and small in magnitude for higher $z$.

\subsection{Determining the amplitude of the kinematic dipole}

We briefly discuss theoretical options for the determination of the dipole amplitude (we do not consider here the issue of optimal estimators and uncertainties). 
We assume that $\bv$ is known from the CMB.

The amplitude of the kinematic dipole as a function of redshift is given by a weighted integral of the boosted correlation function: 
 \be\label{acd}
 { C_1^{\bv}(z,z',m_*)+C_1(z,z',m_*)}=2\pi\int_{-1}^1 d\mu\, \mu\, \la\Delta_{\tilde N}(z,\bn,m<m_*)\,\Delta_{\tilde N}(z',\bn',m'<m_*)\ra,\qquad \mu=\bn\cdot \bn',
 \ee
 which follows from \eqref{cd}, \eqref{cd2}.  The intrinsic dipole is a combination of the dipole from primordial perturbations, which is of the same order as the perturbative $C_\ell \, (\ell>1)$, and the dipole from local structure:{
  \ba
 \tilde C_1 = C_1^{\bv}+C_1^{\rm loc}+ C_1^{\rm prim}.
 \ea 
The perturbative dipole (of order $10^{-5}$) can be neglected relative to the kinematic dipole (of order $10^{-3}$) in the standard cosmology. However, the dipole caused by non-uniform local structure dominates if measurements are made at low redshift. If the dipole is measured at high redshift  ($z\gtrsim 0.7$), the non-uniform local effect is averaged out and the kinematic dipole dominates. 
The local structure dipole is $O(10^{-1})$ for $z\lesssim0.1$, reducing monotonically with $z$, reaching the perturbative level $O(10^{-5})$ in the limit $z\gg1$, so that $ \tilde C_1 \approx C_1^{\bv}$.}
(See \cite{Gibelyou:2012ri} for details.)  

In the case $z'=z$, the upper right panel of Fig.~\ref{fdc} gives examples of $C_1^{\bv}$, in units of {$4\pi v_o^2/9$, or equivalently, of} the CMB dipole amplitude. The results do not hold
at low redshift,  where $C_1^{\bv}$ is not a good approximation to $\tilde C_1$. 

The contribution of $\la\Delta_N\, \Delta_N'\ra$, i.e. the rest-frame correlation function, in \eqref{cd} can be suppressed
if $z'\gg z$ (e.g. $z=0.1, z'=2.5$). The reason is that for widely separated redshifts, the only non-negligible correlations from $\Delta_N$ are those involving the lensing convergence contribution $(5s-2)\kappa$ to $\Delta_N$ [see \eqref{obsn}]. From \eqref{cd}, {
\ba 
\la\Delta_{\tilde N}\,\Delta_{\tilde N}'\ra \approx
 (5s'-2)\la\Delta_N\,\kappa'\ra  +\la\dnv\, \dnv{}' \ra \approx \la\dnv\, \dnv{}' \ra \qquad \mbox{for}\quad z'\gg z,
  \label{ccd} 
\ea
{where the last equality follows since the kinematic dipole is significantly larger than the perturbative quantities.}
The local structure dipole in $\Delta_N$  at the low redshift $z$ dominates over $\dnv$ at $z$ -- but it does not correlate with any of the terms at the high redshift $z'$ and thus we can neglect it. Therefore}
\be \label{ccd2}
\la\Delta_{\tilde N}(z,\bn,m<m_*)\,\Delta_{\tilde N}(z',\bn',m'<m_*)\ra \approx {3\over 4\pi} \,C_1^{\bv}(z,z',m_*)\,\bn\cdot\bn' \qquad \mbox{for}\quad z'\gg z.
\ee
The red curves ($z=0.1$) at $z'=2.5$ in the lower panels of Fig.~\ref{fdc} give examples of $C_1^{\bv}$ (in units of the CMB kinematic dipole amplitude) with $z'\gg z$. 
For intensity mapping, the lensing term in \eqref{obsn}, and hence in \eqref{ccd}, vanishes at first order.

\section{Conclusion}

{Next-generation galaxy surveys over huge volumes of the Universe will deliver the accuracy required to perform a key consistency test of the standard model of cosmology: that the kinematic dipole in the matter distribution should be aligned with that of the CMB. In addition to the direction of the matter kinematic dipole, future surveys will also be able to measure the amplitude of the dipole, which has not previously been investigated for galaxy redshift surveys.}

We derived the expression for the kinematic dipole, due to the velocity $\bv$ of the Solar System observer relative to the rest frame of the CMB, in the number count contrast  of galaxy redshift surveys, including the special case of HI intensity mapping surveys. The kinematic dipole is determined not only by Doppler and aberration effects, but also by redshift and magnitude perturbations. The general expression is given by
\eqref{dipm} and \eqref{dipm2}:
\be\label{2dipm}
\Delta_{\tilde N}=\Delta_N+\dnv,\qquad \dnv(z,\bn,m<m_*)=
{\cal D}(z,m_*)\, \bm{n}\cdot \bvh, \qquad {\cal D}= \left[3+{\dot H \over H^2}+(2-5s){(1+z)\over rH}-b_e\right]v_o.
\ee
For
HI intensity mapping, the magnification bias is effectively $s=2/5$ and $b_e$ is determined by the background brightness temperature. 

The number count contrast observed by the rest-frame observer, $\Delta_N$, is given by the real-space contrast $\delta_N$ plus observational corrections from redshift-space distortions, lensing convergence and other relativistic effects on ultra-large scales.  For the boosted observer, $\Delta_{\tilde N}=\Delta_N+\dnv$, and the boosted correlation function is $\langle\Delta_{\tilde N}\, \Delta_{\tilde N}'\rangle$.
The kinematic dipole term $\dnv$ does not correlate with the terms at the source, $\Delta_N$, and its contribution is much larger than the intrinsic dipole, $\tilde C_1=C_1+C_1^{\bv} \approx C_1^{\bv}$, {provided that we auto-correlate at high enough redshift, or cross-correlate with $z'\gg z$, to beat down the effect of the local structure dipole}. The boosted correlation function and its kinematic dipole are given by \eqref{cd}--\eqref{c1}:
\ba  \label{2cd}
\la\Delta_{\tilde N}\, \Delta_{\tilde N}'\ra 
&=& \la\Delta_{N}\, \Delta_{N}'\ra +\la\dnv\, \dnv{}' \ra
\approx \sum_{\ell=2}{(2\ell+1)\over 4\pi} C_\ell P_\ell + {3\over 4\pi}C_1^{\bv}\,\bn\cdot\bn',
\\
C_1^{\bv}(z,z',m_*)&=& {4\pi\over 9} \,{\cal D}(z,m_*){\cal D}(z',m_*)=2\pi\int_{-1}^1 d\mu\, \mu\, \la\Delta_{\tilde N}(z,\bn,m<m_*)\,\Delta_{\tilde N}(z',\bn',m'<m_*)\ra.
\ea

We presented a theoretical computation of the amplitude of the matter kinematic dipole, using future SKA galaxy and intensity mapping surveys as two examples. The evolution bias and magnification bias for these surveys, based on simulations, are shown in Fig.~\ref{fbs}. The amplitudes ${\cal D}$ and $C_1^{\bv}$ of the kinematic dipoles in these surveys are shown in Fig.~\ref{fdc}, in units of $v_o^2$ and $4\pi v_o^2/9$ (or equivalently, of the corresponding CMB quantity), and as functions of redshift. 

The matter dipole amplitude is larger than that of the CMB for most redshifts -- especially in the case of the HI galaxy survey. In addition, $\dd_{\rm gal}(z,m_*)$ goes through zero and thus $C^\bv_{1\,\rm gal}(z,z,m_*)$ has a minimum of zero, at $z\sim 0.3$. The main driver of this behaviour is the large $b_e$ for galaxies. For angular
correlations across different redshifts, $C_{1\,\rm gal}^\bv(z,z',m_*)$ may decrease or increase monotonically, depending on $z$. For intensity mapping, it is always monotonically decreasing in the redshift range $0<z<3$. 

The amplitude of the matter kinematic dipole can be measured by correlations averaged over directions. When $z'\gg z$, the rest-frame correlations $\langle \Delta_N\,\Delta_N'\rangle$ in \eqref{2cd} are strongly suppressed, with only a small lensing contribution surviving -- which can be neglected relative to the kinematic dipole. 

In principle, measurements of the amplitude give information on the Hubble rate $H(z)$, similarly to the kinematic dipole in luminosity distance \cite{Bonvin:2006en}, shown in \eqref{ddl}. The more complicated kinematic dipole of spectroscopic number counts/ intensity mapping, \eqref{2dipm}, also contains the tracer astrophysical parameters $b_e$ and $s$, which makes extraction of $H$ more difficult -- but the advantage is the high number density. Future work will investigate estimators to measure the amplitude of the kinematic dipole, and the signal-to-noise for various surveys.

\newpage
{\bf Acknowledgments:}\\
We thank Carlos Bengaly and Camille Bonvin for very helpful comments. 
All authors are funded in part by the South African SKA Project and the NRF (South Africa). RM and CC are also supported by the UK STFC, Grants ST/N000668/1 (RM) and  ST/P000592/1 (CC), and SC is also supported by the Claude Leon Foundation.


\begin{thebibliography}{99}

%\cite{Gibelyou:2012ri}
\bibitem{Gibelyou:2012ri} 
  C.~Gibelyou and D.~Huterer,
  ``Dipoles in the Sky,''
  Mon.\ Not.\ Roy.\ Astron.\ Soc.\  {\bf 427}, 1994 (2012)
  %doi:10.1111/j.1365-2966.2012.22032.x
  [arXiv:1205.6476 [astro-ph.CO]].

\bibitem{eb84}
G. F. R. Ellis and J. E.  Baldwin, 
``On the Expected Anisotropy of Radio Source Counts," 
Mon.\ Not.\ Roy.\ Astron.\ Soc.\  {\bf 206}, 377 (1984).

%\cite{Tiwari:2015tba}
\bibitem{Tiwari:2015tba} 
  P.~Tiwari and A.~Nusser,
  ``Revisiting the NVSS number count dipole,''
  JCAP {\bf 1603}, 062 (2016)
  %doi:10.1088/1475-7516/2016/03/062
  [arXiv:1509.02532 [astro-ph.CO]].

%\cite{Bengaly:2016amk}
\bibitem{Bengaly:2016amk} 
C. A. P.  Bengaly,  A.~Bernui, J.~S.~Alcaniz, H.~S.~Xavier and C.~P.~Novaes,
  ``Is there evidence for anomalous dipole anisotropy in the large-scale structure?,''
  Mon.\ Not.\ Roy.\ Astron.\ Soc.\  {\bf 464}, no. 1, 768 (2017)
  %doi:10.1093/mnras/stw2268
  [arXiv:1606.06751 [astro-ph.CO]].

%\cite{Colin:2017juj}
\bibitem{Colin:2017juj} 
  J.~Colin, R.~Mohayaee, M.~Rameez and S.~Sarkar,
  ``High redshift radio galaxies and divergence from the CMB dipole,''
  Mon.\ Not.\ Roy.\ Astron.\ Soc.\  {\bf 471}, 1045 (2017)
  %doi:10.1093/mnras/stx1631
  [arXiv:1703.09376 [astro-ph.CO]].

%\cite{Bengaly:2017zlo}
\bibitem{Bengaly:2017zlo} 
  C.~A.~P.~Bengaly, C.~P.~Novaes, H.~S.~Xavier, M.~Bilicki, A.~Bernui and J.~S.~Alcaniz,
  ``The dipole anisotropy of WISE x SuperCOSMOS number counts,''
  arXiv:1707.08091 [astro-ph.CO].

%\cite{Schwarz:2015pqa}
\bibitem{Schwarz:2015pqa} 
  D.~J.~Schwarz {\it et al.},
  ``Testing foundations of modern cosmology with SKA all-sky surveys,''
  PoS AASKA {\bf 14}, 032 (2015)
  [arXiv:1501.03820 [astro-ph.CO]].

%\cite{Yoon:2015lta}
\bibitem{Yoon:2015lta} 
  M.~Yoon and D.~Huterer,
  ``Kinematic dipole detection with galaxy surveys: forecasts and requirements,''
  Astrophys.\ J.\  {\bf 813}, no. 1, L18 (2015)
  %doi:10.1088/2041-8205/813/1/L18
  [arXiv:1509.05374 [astro-ph.CO]].

%\cite{Ghosh:2016tbj}
\bibitem{Ghosh:2016tbj} 
  S.~Ghosh, P.~Jain, G.~Kashyap, R.~Kothari, S.~Nadkarni-Ghosh and P.~Tiwari,
  ``Probing statistical isotropy of cosmological radio sources using SKA,''
  J.\ Astrophys.\ Astron.\  {\bf 37}, 25 (2016)
  %doi:10.1007/s12036-016-9395-8
  [arXiv:1610.08176 [astro-ph.CO]].

%\cite{Challinor:2011bk}
\bibitem{Challinor:2011bk} 
  A.~Challinor and A.~Lewis,
 ``The linear power spectrum of observed source number counts,''
  Phys.\ Rev.\ D {\bf 84}, 043516 (2011)
  [arXiv:1105.5292].

%\cite{Jeong:2011as}
\bibitem{Jeong:2011as} 
  D.~Jeong, F.~Schmidt and C.~M.~Hirata,
  ``Large-scale clustering of galaxies in general relativity,''
  Phys.\ Rev.\ D {\bf 85}, 023504 (2012)
 % doi:10.1103/PhysRevD.85.023504
  [arXiv:1107.5427 [astro-ph.CO]].

%\cite{Alonso:2015uua}
\bibitem{Alonso:2015uua} 
  D.~Alonso, P.~Bull, P.~G.~Ferreira, R.~Maartens and M.~Santos,
  ``Ultra large-scale cosmology in next-generation experiments with single tracers,''
  Astrophys.\ J.\  {\bf 814}, no. 2, 145 (2015)
  %doi:10.1088/0004-637X/814/2/145
  [arXiv:1505.07596 [astro-ph.CO]].

%\cite{Bonvin:2005ps}
\bibitem{Bonvin:2005ps} 
  C.~Bonvin, R.~Durrer and M.~A.~Gasparini,
  ``Fluctuations of the luminosity distance,''
  Phys.\ Rev.\ D {\bf 73}, 023523 (2006)
  Erratum: [Phys.\ Rev.\ D {\bf 85}, 029901 (2012)]
 % doi:10.1103/PhysRevD.85.029901, 10.1103/PhysRevD.73.023523
  [astro-ph/0511183].

%\cite{Hall:2012wd}
\bibitem{Hall:2012wd} 
  A.~Hall, C.~Bonvin and A.~Challinor,
  ``Testing General Relativity with 21-cm intensity mapping,''
  Phys.\ Rev.\ D {\bf 87}, no. 6, 064026 (2013)
 % doi:10.1103/PhysRevD.87.064026
  [arXiv:1212.0728 [astro-ph.CO]].

%\cite{Slosar:2016utd}
\bibitem{Slosar:2016utd} 
  A.~Slosar,
  ``Dipole of the Epoch of Reionization 21-cm signal,''
  Phys.\ Rev.\ Lett.\  {\bf 118}, no. 15, 151301 (2017)
 % doi:10.1103/PhysRevLett.118.151301
  [arXiv:1609.08572 [astro-ph.CO]].

%\cite{Aghanim:2013suk}
\bibitem{Aghanim:2013suk} 
  N.~Aghanim {\it et al.} [Planck Collaboration],
  ``Planck 2013 results. XXVII. Doppler boosting of the CMB: Eppur si muove,''
  Astron.\ Astrophys.\  {\bf 571}, A27 (2014)
  [arXiv:1303.5087].

%\cite{Maartens:2015mra}
\bibitem{Maartens:2015mra} 
  R.~Maartens {\it et al.} [SKA Cosmology SWG Collaboration],
  ``Overview of Cosmology with the SKA,''
  PoS AASKA {\bf 14}, 016 (2015)
  [arXiv:1501.04076 [astro-ph.CO]].

%\cite{Camera:2014bwa}
\bibitem{Camera:2014bwa} 
  S.~Camera, M.~G.~Santos and R.~Maartens,
  ``Probing primordial non-Gaussianity with SKA galaxy redshift surveys: a fully relativistic analysis,''
  Mon.\ Not.\ Roy.\ Astron.\ Soc.\  {\bf 448}, no. 2, 1035 (2015) 
  [Erratum: Ibid., {\bf 467}, 1505 (2017)]
  %doi:10.1093/mnras/stv040
  [arXiv:1409.8286 [astro-ph.CO]].

%\cite{Santos:2015gra}
\bibitem{Santos:2015gra} 
  M.~G.~Santos {\it et al.},
  ``Cosmology with a SKA HI intensity mapping survey,''
  PoS AASKA14 (2015) 019
  [arXiv:1501.03989 [astro-ph.CO]].

%\cite{Bonvin:2006en}
\bibitem{Bonvin:2006en} 
  C.~Bonvin, R.~Durrer and M.~Kunz,
  ``The dipole of the luminosity distance: a direct measure of H(z),''
  Phys.\ Rev.\ Lett.\  {\bf 96}, 191302 (2006)
 % doi:10.1103/PhysRevLett.96.191302
  [astro-ph/0603240].

\end{thebibliography}
\end{document}